\documentclass[twocolumn,showpacs,preprintnumbers,amsmath,amssymb,prb,floatfix]{revtex4}

\usepackage{graphicx}% Include figure files
\usepackage{dcolumn}% Align table columns on decimal point 
\usepackage{bm}% bold math
\usepackage{epstopdf} 

\begin{document}
\bibliographystyle{prsty}

\title{Effects of Strain on Orbital Ordering and Magnetism at the Perovskite Oxide Interfaces: LaMnO$_3$/SrMnO$_3$
}% Force line breaks with \\

\author{B. R. K. Nanda}
\author{Sashi Satpathy}%
\affiliation{% 
Department of Physics $\&$ Astronomy, University of Missouri,
Columbia, MO 65211}%

\date{\today}% It is always \today, today,

\begin{abstract}
 We study how strain affects orbital ordering and magnetism at the interface
 between SrMnO$_3$ and LaMnO$_3$ from density-functional calculations and interpret the basic results in terms of a three-site Mn-O-Mn model. 
 Magnetic interaction between the Mn atoms is governed by a competition between the antiferromagnetic superexchange of the Mn t$_{2g}$ core spins and the ferromagnetic double exchange of the itinerant e$_g$ electrons. While the core electrons are relatively unaffected by the strain, the orbital character of the itinerant electron is strongly affected, which in turn causes a large change in the strength of the ferromagnetic double exchange.  The epitaxial strain produces the tetragonal distortion of the MnO$_6$ octahedron, splitting the
Mn-e$_g$ states into x$^2$-y$^2$ and 3z$^2$-1 states, with the former being lower in energy, if the strain is tensile in the plane, and opposite if the strain is compressive. For the case of the tensile strain, the resulting higher
occupancy of the x$^2$-y$^2$ orbital enhances the in-plane ferromagnetic
double exchange owing to the larger electron hopping in the plane, causing at the same time
a reduction of the out-of-plane double exchange. This reduction is large enough
to be overcome by antiferromagnetic superexchange, which wins to produce a net antiferromagnetic interaction between the out-of-plane Mn atoms.
For the case of the in-plane compressive  strain, the reverse happens, viz., that the higher occupancy of the
3z$^2$-1 orbital results in the out-of-plane ferromagnetic interaction, while the in-plane magnetic interaction remains antiferromagnetic. Concrete density-functional results are presented for the (LaMnO$_3$)$_1$/(SrMnO$_3$)$_1$ and (LaMnO$_3$)$_1$/(SrMnO$_3$)$_3$ superlattices for various strain conditions.
\end{abstract}

\maketitle

\section{Introduction}

\begin{figure}
\includegraphics[width=9.0cm]{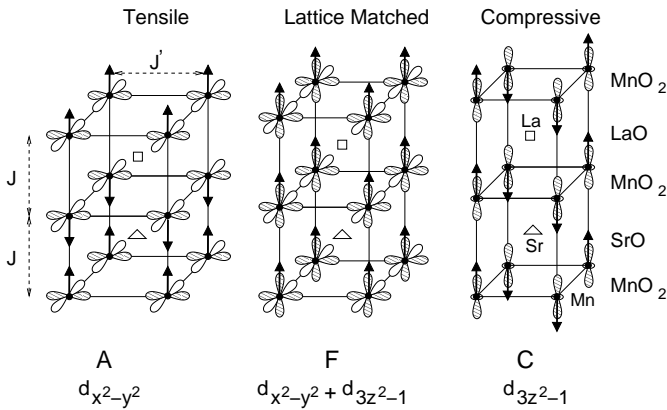}
\caption{\label{magfig} Different magnetic configurations considered in this paper for the (LMO)$_1$/(SMO)$_1$ superlattice. `A' stands for the structure with ferromagnetic ordering in the MnO$_2$ plane and antiferromagnetic ordering between the planes, while `F' stands for ferromagnetic ordering in all directions and `C' stands for antiferromagnetic ordering in the MnO$_2$ plane and  ferromagnetic ordering between the planes. The schematic orbital ordering shown in the figure was found from our density functional results presented below and was also inferred from the experiments\cite{yamada}. The symbols J and J$^{\prime}$ denote, respectively, the out-of-plane and  in-plane exchange interactions between the Mn atoms. The strain condition under which each structure is stabilized has been indicated in the figure. The oxygen atoms which occur at the mid-point between two neighboring Mn atoms have not been shown.}
\end{figure}

Recent advances in successfully designing atomically sharp interfaces between dissimilar transition metal oxides  have revealed the formation of new electronic and magnetic phases 
 at the vicinity of the interface, which are qualitatively different from the parent compounds. 
The interfacial phases show diverse magnetic properties due to the coupling between charge, orbital and spin degrees of freedom.
 For example the magnetic ordering at the interface between the two antiferromagnetic insulators SrMnO$_3$ (G-type) and LaMnO$_3$ (A-type), schematically shown in Fig. \ref{magfig}, 
 could be ferromagnetic along all directions, ferromagnetic in the xy-plane and antiferromagnetic normal to the plane, or antiferromagnetic in the plane and ferromagnetic normal to the plane depending on the composition of the parent compounds and epitaxial strain on the interface \cite{yamada, koida, salva, anand1, anand2, anand3}. 

   The epitaxial strain, arising due to lattice mismatch between the constituent compounds of the superlattice and the substrate, induces anisotropic hopping between orbitals to cause orbital ordering at the interface. 
By varying the strain condition the orbital ordering changes which in turn changes the magnetic ordering at the interface. In this paper we examine the magnetic properties at the interface of SrMnO$_3$ (SMO) and LaMnO$_3$ (LMO) for different epitaxial strain conditions through first principles electronic structure calculations.

  Experimental studies show that
if the substrate induces tensile strain at the interface of the LMO/SMO superlattice, where the in-plane lattice parameter `a' is greater than the out-of-plane lattice parameter `c', as in the case of (LMO)$_3$/(SMO)$_2$ superlattice grown on SrTiO$_3$ (STO) substrate, the magnetic ordering of the interfacial Mn atoms is A-type 
 with in-plane (MnO$_2$ plane) ferromagnetic (FM) ordering and out-of-plane (between MnO$_2$ planes) antiferromagnetic (AFM) ordering \cite{yamada}. 
   
 Quite interestingly, when the (LMO)$_3$/(SMO)$_2$ superlattice is grown on La$_{0.3}$Sr$_{0.7}$Al$_{0.65}$Ta$_{0.35}$O$_3$ (LSAT) substrate, which induces no strain (a $\sim$ c), the interface shows a three dimensional FM ordering (F-type) \cite{yamada}. If the interface experiences a
compressive strain (a $<$ c), as in the case of LMO/SMO superlattice grown on 
LaAlO$_3$ (LAO) substrate, the magnetic ordering is C-type with in-plane AFM ordering and out-of-plane FM ordering \cite{yamada}. 

\begin{figure}
\includegraphics[width=8.5cm]{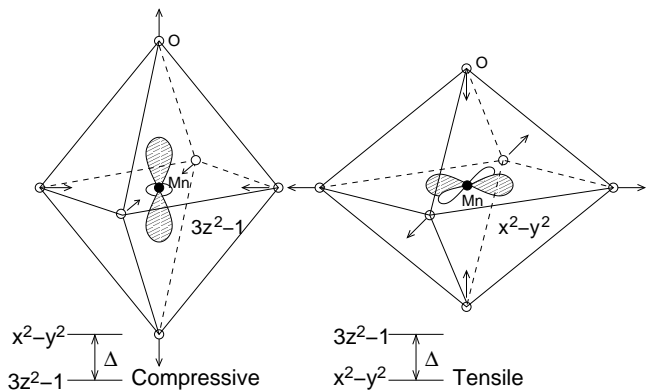}
\caption{\label{straindemo} Energy splitting of the Mn(e$_g$) orbitals at the LMO/SMO interface for compressive and tensile strain conditions. The parameter $\Delta$ is the difference between the energies of the Mn-d$_{x^2-y^2}$ and the Mn-d$_{3z^2-1}$ orbitals. Compressive strain makes the 3z$^2$-1 orbital lower in energy while tensile strain makes it higher.}
\end{figure}

  Substrates are instrumental in inducing epitaxial strain and thereby enforce tetragonal distortion to the
superlattice. As a consequence, in case of LMO/SMO superlattice, the substrate distorts the MnO$_6$ octahedron and  splits the degenerate Mn-e$_g$ states into x$^2$-y$^2$ and 3z$^2$-1 states. Varied tetragonal distortion changes the onsite energy and hence the occupancy of these two non-degenerate e$_g$ states (Fig. \ref{straindemo}). Since the electronic configuration of  Mn atoms away from the interface is  the same as in the bulk compounds, (Mn$^{4+}$, t$_{2g}^{3}$e$_{g}^0$) for SMO and (Mn$^{3+}$, t$_{2g}^{3}$e$_{g}^1$) for LMO, strain is not expected to affect the magnetic configuration of the inner MnO$_2$ layers to a large extent. However, at the interface, where we see the valence state of the Mn atoms lies between 3+ and 4+ because of charge reconstruction \cite{koida, satoh}, the varied occupancy of the non-degenerate e$_g$ orbitals imposes different orbital ordering for different strain conditions (Fig. \ref{straindemo}) and influences the interface magnetism considerably.


 In this paper, we have studied in detail the interfacial magnetic properties of LMO/SMO superlattices for different strain conditions by performing electronic
structure calculations based on the density functional theory. To illustrate the strain effect on magnetism, we have proposed a simple three-site model to calculate
 the interfacial Mn-O-Mn magnetic exchange both in the MnO$_2$ plane and between the planes for different strain conditions. From the model we see that the onsite energy difference between x$^2$-y$^2$ and 3z$^2$-1 orbitals (Fig. \ref{straindemo}) is instrumental in switching the ferromagnetic and antiferromagnetic interactions. When the 3z$^2$-1 orbital is sufficiently lower in energy than the x$^2$-y$^2$ orbital (compressive strain), the Mn-O-Mn exchange is antiferromagnetic in the plane and ferromagnetic between the planes, and opposite when x$^2$-y$^2$ orbital is sufficiently lower in energy (tensile strain). If the energy levels of both the e$_g$ orbitals are close enough (lattice matched interface), then the Mn-O-Mn exchange is ferromagnetic in all directions. 

       The rest of the paper is organized as follows. In section II we describe the structural and computational details. A detailed analysis of the electronic structure of the (LMO)$_1$/(SMO)$_1$ superlattice at different strain conditions, obtained from the density-functional calculations, is carried out in section III. In section IV, we illustrate the effect of epitaxial strain on the magnetic ordering, with the aid of a proposed three site (Mn-O-Mn) model. Electronic and magnetic properties of the (LMO)$_1$/(SMO)$_3$ superlattice at different strain conditions are discussed in section V. Finally in section VI we present the summary.    

\section {Structural and Computational Details}
  We have taken the equivalent cubic perovskite structure of LMO and SMO in order to study the electronic and magnetic properties at the interface of these two compounds with the aid of first principles electronic structure calculations. The effect of epitaxial strain, which arises due to lattice mismatch between the substrate and the LMO/SMO superlattice, is taken into account
by applying tetragonal distortion to the superlattice.

    The tetragonal distortion is quantified by the `c/a' ratio which differs from one. Here `a' is the in-plane (xy-plane) lattice parameter which coincides with the lattice parameter of the substrate and `c' is the average out-of-plane lattice parameter (along z-axis). The `c/a' ratio is determined from the linear relation: 
c - a$_0$ = -4$\nu$(a - a$_0$), where `a$_0$' is the in-plane lattice parameter of the superlattice when there is no strain (c/a = 1) and coefficient $\nu$ is the Poisson's ratio which is approximately 0.3 for perovskite manganites \cite{yamada, buch}. Experimentally it is found that for LMO/SMO superlattices, a$_0$ matches with the weighed average of the lattice constants of bulk LMO (3.936 \AA) and bulk SMO (3.806 \AA) \cite{yamada}. For example, for (LMO)$_1$/(SMO)$_1$ superlattice, the value of a$_0$ is $\frac{1}{2}$(3.936 + 3.806) \AA.

  In this paper, we have considered two 
superlattices, viz., (LMO)$_1$/(SMO)$_1$ and (LMO)$_1$/(SMO)$_3$ to study the electronic and magnetic properties at different strain conditions. As is well-known, the strength of the Jahn-Teller distortion is less  in the mixed   
compounds (La, Sr)MnO$_3$ as compared to that of LaMnO$_3$, we have considered  a small Jahn-Teller distortion (Q$_2 \approx$\ 0.05 \AA) in the basal plane for the interfacial MnO$_2$ layers. However, test calculations showed that a small variation of Q$_2$ does not change the electronic and magnetic properties of the superlattice qualitatively.

   All electronic structure calculations reported in this work have been performed using the self-consistent tight-binding linearized muffin-tin orbital (TB-LMTO)  method with 
 the atomic sphere approximations (ASA) \cite{lmto}. Self-consistent calculations are done within the framework of generalized gradient 
approximation including Coulomb correction (GGA+U). All results are obtained with U = 5 eV and J = 1 eV unless otherwise stated.

\section{Electronic structure of the $(LaMnO_3)_1/(SrMnO_3)_1$ superlattice}

 In this section, we describe the effect of strain on the electronic structure at the interface from {\it ab initio} density-functional (DFT) 
calculations. We focus on the (LMO)$_1$/(SMO)$_1$ superlattice and our results suggest that many of the interfacial electronic and magnetic properties shown by 
this superlattice should also be valid for the more general (LMO)$_n$/(SMO)$_m$ superlattices.  

   We briefly summarize the electronic structure and magnetism for the bulk SMO and LMO compounds. In bulk SMO, the Mn atoms are in 4+ charge state to have  
three d-electrons which are occupied in the triply degenerate t$_{2g}$ states. The doubly degenerate e$_g$ states, which are higher in energy with respect to t$_{2g}$ states because of a MnO$_6$  octahedral crystal field split, remain unoccupied. The t$_{2g}^3$ spin majority states mediate an antiferromagnetic superexchange to stabilize the G-type antiferromagnetic ordering in the bulk SMO compound, where spin of each Mn atom is opposite to that of the nearest neighbor Mn atoms \cite{milis, goodenough}.

     In bulk LMO the Mn atoms are in 3+ charge state to have four d-electrons. Three electrons are occupied in the localized t$_{2g}$ states and the remaining one electron is occupied in the e$_g$ state. The Jahn-Teller (JT) distortion to the MnO$_6$ octahedron further splits the e$_g$ states into two non-degenerate states: e$_{g}^1$ which is lower in energy and e$_{g}^2$ which is higher in energy \cite{sashi1}. The one e$_g$ electron is occupied in the e$_{g}^1$ state whose lobes are pointed towards the longest Mn-O bond. The JT distortion stabilizes the A-type antiferromagnetic structure in the LMO compound \cite{feinberg}.

  At the LMO/SMO interface the Mn atoms do not satisfy the 4+ charge state or the 3+ charge state to support the bulk magnetism of SMO or LMO. The mixed valence nature of the Mn atoms as well as the effect of epitaxial 
 strain create diverse magnetic phases at the interface, which will be analyzed in this section.

\begin{figure}
\includegraphics[width=6.0cm]{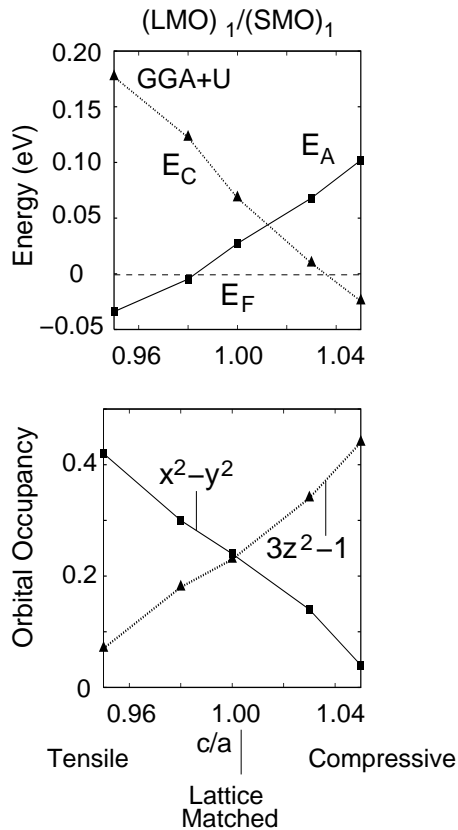}
\caption{\label{energyoccup} Total energies for the magnetic configurations, A and C, relative to the energy for the configuration F, as a function of the tetragonal distortion c/a (top).  The magnetic configurations A, F, and C are shown in Fig. \ref{magfig}. Bottom figure shows the occupancies of the x$^2$-y$^2$ and the 3z$^2$-1 orbitals per Mn atom as a function of the tetragonal distortion.}
\end{figure}

  Epitaxial strain, arising due to the substrate on which the interface is grown, induces tetragonal distortion to the cubic interface which is quantified by the `c/a' ratio that differs from one. Experimental studies show different magnetic behavior at the interface for different `c/a' ratios \cite{yamada}. To obtain the dependence of the magnetic ground state on the strain condition, we have performed total energy calculations in the range 
0.95 $\le$ c/a $\le$ 1.05 for three possible magnetic configurations (A, F, and C) (Fig. \ref{magfig}). Magnetic configuration A represents the FM ordering
 in the MnO$_2$ plane and AFM ordering between the 
planes. Magnetic configuration C represents the AFM ordering in the MnO$_2$ 
plane and  FM ordering between the planes and F represents the FM ordering in all 
directions. The energetics are shown in Fig. \ref{energyoccup} (top). 

     From the figure we see that for a strong compressive strain ({\it e.g.} c/a = 0.95), `A' is the most stable magnetic configuration. For the lattice matched structure ( c/a = 1, no strain), the interface stabilizes with magnetic configuration F and in case of a strong compressive strain ({\it e.g.} c/a = 1.05) it stabilizes with magnetic configuration C. 
The results are in accordance with the experimental observations
 which show that when the substrates are STO (c/a = 0.98), LSAT (c/a = 1.01), and 
LAO (c/a = 1.05), the respective magnetic configurations at the LMO/SMO interface are A, F, and C.\cite{yamada}
      
      We see that as strain changes, the occupancy of the e$_g$ orbitals,  which controls the magnetic interaction at the interface, also changes. This is shown in Fig. \ref{energyoccup} (bottom). For the tensile strain condition (c/a $<$ 1) the occupancy of the x$^2$-y$^2$ orbital is greater than the
occupancy of the 3z$^2$-1 orbital and opposite if the strain is compressive (c/a $>$ 1). For the lattice matched structure  (c/a = 1) both the e$_g$ orbitals are more or less equally occupied. Fig. \ref{energyoccup} also shows that for any value of `c/a', the non-degenerate e$_g$ states combinedly occupy 0.5 electrons which along with three t$_{2g}$ core electrons make the average valence of the interface Mn atoms to be +3.5 as expected. 

 Magnetic interaction between the Mn atoms is determined by the competition  between ferromagnetic double exchange \cite{hase, zener,  genes} via the itinerant Mn-e$_g$ electrons and antiferromagnetic superexchange between the localized Mn-t$_{2g}$ core spins. When  x$^2$-y$^2$ is more occupied and 3z$^2$-1 orbital is less occupied (or unoccupied), the strong double exchange in the MnO$_2$ plane strengthens the ferromagnetic ordering while superexchange stabilizes the antiferromagnetic ordering between the planes. The magnetic ordering is opposite to the above when the occupancies of the
 two e$_g$ orbitals are reversed. If both the e$_g$ orbitals are more or less equally
 occupied, the double exchange stabilizes the ferromagnetic ordering both in the plane and between the planes.  

As described in the following subsections, a detailed analysis of the density-functional electronic structure 
 of the LMO/SMO interface under different strain conditions
gives us a better understanding on the strain induced orbital ordering and its effect on magnetic properties at the interface.

\subsection{c/a = 0.95, Tensile strain}

\begin{figure}
\includegraphics[width=7.7cm]{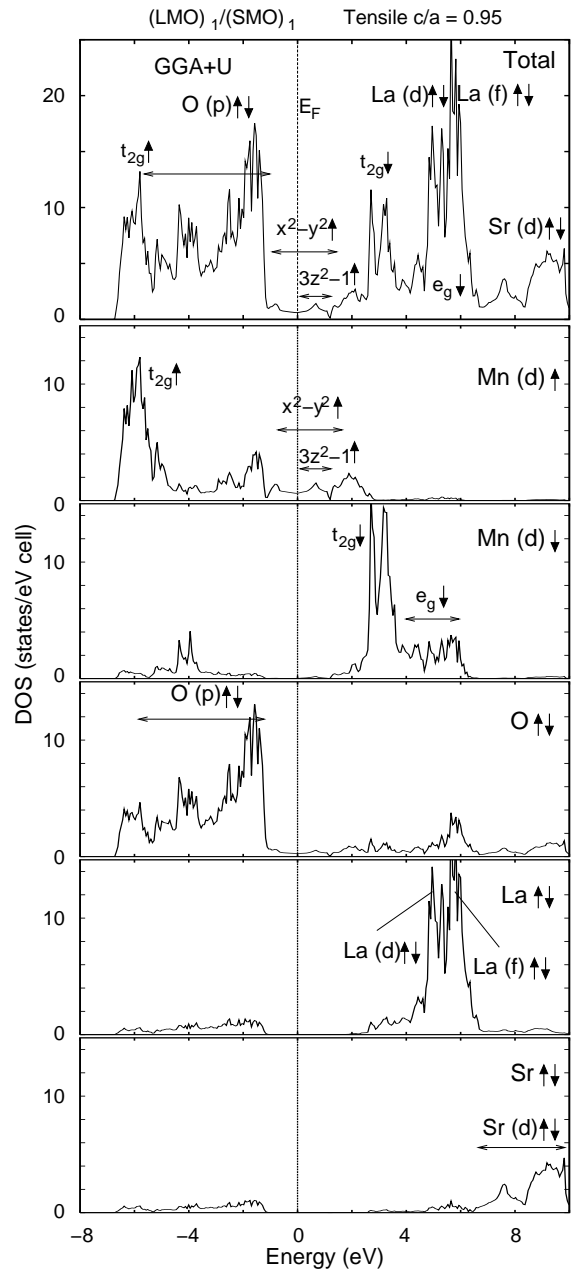}
\caption{\label{dosaaf}Total and partial DOS for the (LMO)$_1$/(SMO)$_1$ superlattice (c/a = 0.95) in the A-type magnetic configuration. The symbols $\uparrow$ and $\downarrow$ represent the local spin of the atoms. The Mn-e$_g$$\uparrow$ state at the Fermi level (E$_F$) splits into x$^2$-y$^2$  and 3z$^2$-1 states. The orbital character of the e$_g$ states at E$_F$ is shown in Fig. \ref{band}.}
\end{figure}

Tensile strain reduces the out-of-plane lattice parameter c and enhances the in-plane lattice parameter a. In other
words it decreases the Mn-O bond length between the MnO$_2$ planes and increases it in the plane. In such a scenario, the total energy calculation (Fig. \ref{energyoccup}) suggests a stable A-type magnetic configuration (Fig. \ref{magfig}) when tetragonal distortion (c/a) is close to 0.95.  In Fig. \ref{dosaaf}, we have shown the total and partial densities of states (DOS)  for the (LMO)$_1$/(SMO)$_1$ superlattice (c/a = 0.95) in the A-type structure obtained from the GGA+U calculations.

\begin{figure}
\includegraphics[width=8.5cm]{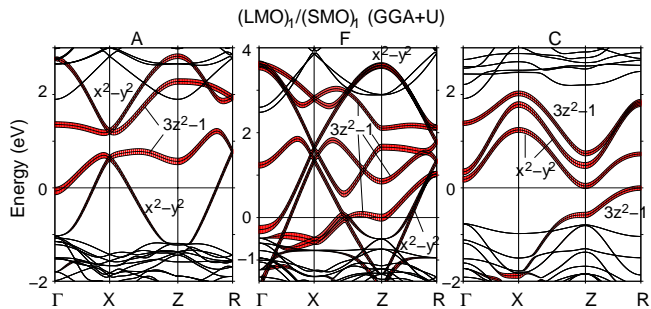}
\caption{\label{band}
 (color online). Orbital character of the electron bands near E$_F$ for the three magnetic structures A, F, and C. The bands are plotted along the high symmetry points 
$\Gamma$ (0, 0, 0), X ($\frac{\pi}{2a}$, -$\frac{\pi}{2a}$, 0), Z (0, 0, -$\frac{\pi}{2c}$), and R ($\frac{\pi}{2a}$, -$\frac{\pi}{2a}$, -$\frac{\pi}{2c}$). The unit cell for the magnetic structures is doubled along the xy-plane with the formula unit 2 $\times$ (LMO)$_1$/(SMO)$_1$. For the AFM configurations (A and C), we have two Mn$\uparrow$ and two Mn$\downarrow$ atoms. Only spin-up bands are shown for the FM configuration (F).}
\end{figure}

   The characteristic features of the electronic structure under tensile strain as seen from Fig. \ref{dosaaf} are as follows. The localized Mn-t$_{2g}$ states lie far below the Fermi level (E$_{F}$) because of the octahedral crystal field and strong Coulomb repulsion. The  O-p states occur in the energy range of  -6 to -1 eV. The x$^2$-y$^2$ and 3z$^2$-1 orbitals are
 predominant at E$_{F}$. Since the intraplane (on the xy-plane) Mn-O bond is longer than the interplane (along the z-axis) one (Fig. \ref{straindemo}), this lowers the energy of the x$^2$-y$^2$ orbital making it more occupied and raises the energy of the 3z$^2$-1 orbital, which becomes 
less occupied.

\begin{figure}
\includegraphics[width=6cm]{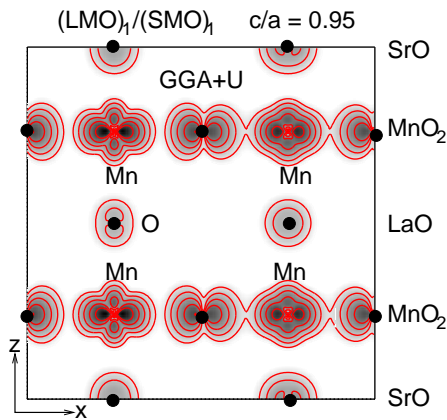}
\caption{\label{contouraaf} (color online). Valence electron charge-density contours plotted
on the xz-plane in the energy range E$_F$ - 0.15 eV to E$_F$ to indicate the
orbital ordering for the A-type magnetic configuration. Contour values are: $\rho_n$ = $\rho_0 \times 10^{n\delta}$e/\AA$^3$, where $\rho_0$ = 3.7 $\times 10^{-3}$, $\delta$ = 0.4 and n labels the contours. The charge contours on the yz-plane (not shown) are identical to that of xz-plane. The orbital ordering is mainly x$^2$-y$^2$.}
\end{figure}

The origin behind the stability of A-type magnetic configuration for the tensile interface is explained below. In the bulk LMO, Mn (3+) atom has the electronic configuration t$_{2g}^3$e$_{g}^1$ and in the bulk SMO, Mn (4+) atom has the electronic configuration t$_{2g}^3$e$_{g}^0$. Since at the interface,  the MnO$_2$ layers are surrounded by (SrO)$^0$ layer and (LaO)$^{1+}$ layer, the interface Mn atoms are left with the average valence state of +3.5. In such a scenario, the t$_{2g}$ orbitals
 will occupy 3 electrons in the spin majority states and the e$_g$ orbitals will occupy the remaining 0.5 electrons.

     Without any occupancy of the e$_g$ states, the only
contribution to the energy comes from the superexchange  interaction between the localized t$_{2g}$ states to stabilize the G-type AFM phase as in the case of SMO. However, the itinerant e$_g$ states, if partially
 occupied, can mediate the Anderson-Hasegawa double exchange\cite{hase} to stabilize the FM phase.

    The strength of the FM ordering in the plane or out of the plane
depends on the occupancy of the individual x$^2$-y$^2$ and 3z$^2$-1 orbitals.
 From our calculations
(Fig. \ref{energyoccup}) we find that for tensile strain condition (c/a = 0.95), the occupancy of  x$^2$-y$^2$  orbital is close to 0.45, while for   3z$^2$-1  orbital it is less than  0.1. This is also reflected in the Mn-e$_g$ band dispersion for the A-type
magnetic configuration shown in Fig. \ref{band}. Since, the unit cell for the magnetic structure is doubled along the xy-plane (i.e. 2$\times$ (LMO)$_1$/(SMO)$_1$), for the AFM magnetic configuration we have two Mn$\uparrow$ and two Mn$\downarrow$ atoms.  Hence, for the local spin majority channel, we have two x$^2$-y$^2$ orbitals and two 3z$^2$-1 orbitals. From the figure we see that the 3z$^2$-1 orbitals are mostly in the conduction band and only one x$^2$-y$^2$ orbital of two crosses the Fermi level and lies mostly in the valence band. This implies that almost one electron per two Mn atoms in the x$^2$-y$^2$ states is occupied which is consistent with the orbital occupancy calculation.

  In such a case the  x$^2$-y$^2$ orbitals will mediate the double exchange mechanism in the MnO$_2$ plane to
stabilize a FM ordering in the plane.  The gain in kinetic energy due to the planar orbital order, induced by the anisotropic hopping, is more than the loss of super exchange energy. Since the 3z$^2$-1 orbitals are only marginally occupied, superexchange between the localized t$_{2g}$ electrons stabilizes the AFM ordering between the MnO$_2$ planes. The net result is an A-type AFM ordering at the interface.

    The valence electron charge density contours for states in the vicinity of the Fermi level (E$_F$), shown in Fig. \ref{contouraaf}, provides a visualization
of the above analysis. The charge contours show that  the orbital ordering is predominantly Mn-x$^2$-y$^2$, O-p$_x$ and p$_y$, while the occupancies of the 3z$^2$-1 and p$_z$ orbitals are small. As a result we see a strong coupling between the Mn-e$_g$ and O-p orbitals in the plane while it is rather weak between the planes. Therefore, the in-plane  magnetic exchange interaction J$^{\prime}$ is ferromagnetic while the out-of-plane J is antiferromagnetic (Fig. \ref{magfig}).
Our results are consistent with the experimental results that the magnetic ordering at the 
interface for (LMO)$_3$/(SMO)$_2$ superlattice grown on STO substrate (c/a = 0.98) is A-type \cite{yamada}.

\subsection {c/a=1.0, Lattice matched structure}

\begin{figure}
\includegraphics[width=9.0cm]{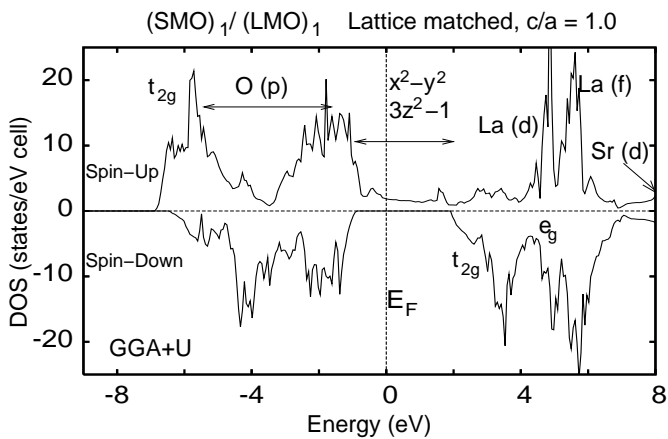}
\caption{\label{dosfm}Total spin-up and spin-down DOS for (LMO)$_1$/(SMO)$_1$ superlattice in the F-type magnetic configuration. Both x$^2$-y$^2$ and 3z$^2$-1 orbitals are more or less equally occupied. The orbital character of the e$_g$ states at E$_F$ in the spin-up channel is shown in Fig. \ref{band}. The superlattice shows the half-metallic behavior.}
\end{figure}

\begin{figure}
\includegraphics[width=6cm]{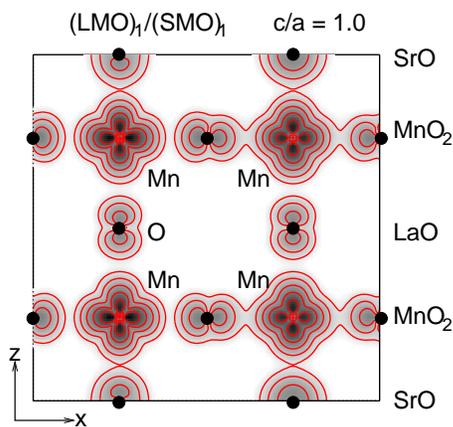}
\caption{\label{contourfm} (color online). Valence electron charge-density contours plotted
on the xz-plane in the energy range E$_F$ - 0.15 eV to E$_F$ to indicate the
orbital ordering for the F-type magnetic configuration. Contour values are: $\rho_n$ = $\rho_0 \times 10^{n\delta}$e/\AA$^3$, where $\rho_0$ = 3.7 $\times 10^{-3}$, $\delta$ = 0.4 and n labels the contours. The charge contours on the yz-plane (not shown) are identical to that of xz-plane. The orbital state of each Mn is a mixture of x$^2$-y$^2$ and 3z$^2$-1.}
\end{figure}

        Lattice matched interfaces are without any tetragonal distortion and hence the in-plane and out-of-plane Mn-O bond lengths are identical. Total energy calculation (Fig. \ref{energyoccup}) in this case favors a three dimensional FM ordering (F-type). To gain insight into the origin behind the FM ground state, we analyze the electronic structure for the lattice matched interface. In Fig. \ref{dosfm}, we have shown the total spin-up and spin-down DOS for the F-type magnetic configuration. 

  General features of the electronic structure of the lattice matched interface are 
 similar to that of the tensile interface. However, now on either side of the the Fermi level, both x$^2$-y$^2$ and 3z$^2$-1 orbitals are predominant in the spin-up channel and they have nearly equal onsite energies. It is due to the fact that the Mn-O bond lengths are same both in-plane and  out-of-plane, making the e$_g$ states nearly degenerate in energy.  This is also substantiated from the dispersion of the spin-up Mn-e$_g$ bands for the F-type structure shown in Fig. \ref{band}. Since the formula unit is doubled along the xy-plane, there are four Mn atoms and all are in the same spin orientation. Hence, in the spin-up channel, we have eight e$_g$ bands of which almost six lie in the conduction band. Of the remaining two bands, which are part of the valence bands, one is predominantly of 3z$^2$-1 character, while the other is predominantly x$^2$-y$^2$. Hence the occupancy of each of these orbitals is close to a quarter electron per Mn atom which is also seen from the orbital occupancy results of  Fig. \ref{energyoccup}.  The valence charge-density contours of Fig. \ref{contourfm} indicates the orbital occupancy of the two Mn-e$_g$ orbitals as well as their hybridization with the O-p orbitals.   

   The  partially occupied x$^2$-y$^2$ and 3z$^2$-1 orbitals mediate a ferromagnetic double exchange, strong enough to overcome the antiferromagnetic superexchange both in the plane and out of the plane to stabilize a three dimensional FM ordering. We have shown earlier\cite{ranjit} that in the case of CaMnO$_3$/CaRuO$_3$ interface, a leaking of 0.2 electrons from the metallic
   CaRuO$_3$ side  to the Mn-e$_g$ states near the interface, which were otherwise unoccupied, is sufficient to stabilize the FM ordering of the Mn spins. In the present case, both the e$_g$ orbitals being occupied substantially  (more than 0.2 electrons in each orbital), a strong ferromagnetic
double exchange coupling along all directions is expected. This is consistent with the experimental observation of ferromagnetism  in the LMO/SMO interface structures grown on the LSAT substrate (c/a = 1.01) \cite{yamada}. 
%Similarly, a strain free film of La$_{0.5}$Sr$_{0.5}$MnO$_3$\cite{tokurafilm}, whose electronic configuration is identical to this interface,  also stabilizes in the FM ground state.

         The other prominent feature in the electronic structure of the lattice matched interface is the opening of a gap
at the Fermi level in the spin-down channel which makes the system half-metallic (Fig. \ref{dosfm}). 
%We note that bulk La$_{1-x}$Sr$_x$MnO$_3$ (0.3 $\leq$ x $\geq$ 0.7) shows half-metallic behavior \cite{venkat}, which arises due to the strong exchange splitting of the Mn-d states. Fig. \ref{dosfm} shows that the Mn-d states lie above the Fermi level in the spin-down channel to open a gap.  In the spin-up channel we see that the Mn-t$_{2g}$ states are completely occupied and lie far below the Fermi level. The partial occupancy of the itinerant Mn-e$_g$ states as discussed above makes the Fermi level occupied in the spin-up channel to make the (LMO)$_1$/(SMO)$_1$ superlattice half-metallic. 

\subsection{c/a = 1.05, Compressive strain}

\begin{figure}
\includegraphics[width=6.5cm]{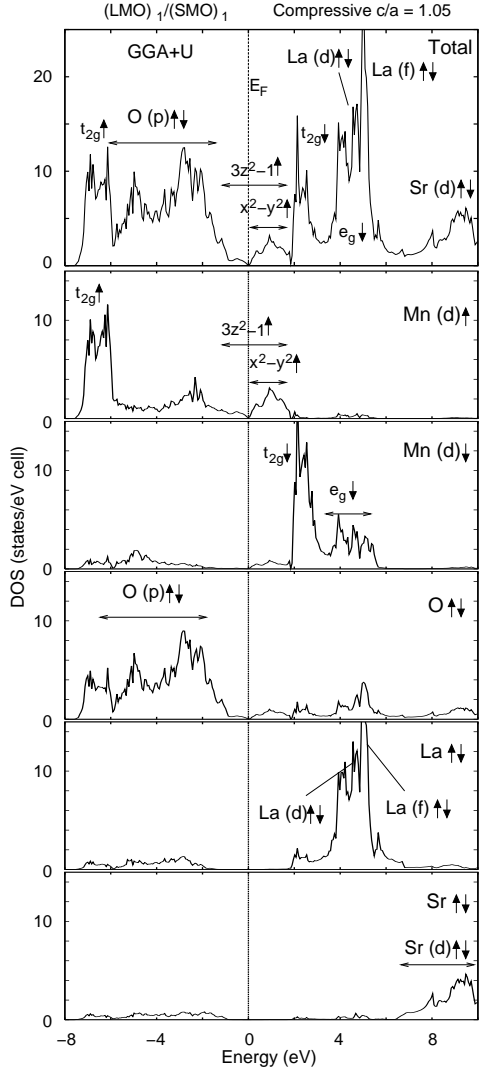}
\caption{\label{doscaf}Total and partial DOS for the (LMO)$_1$/(SMO)$_1$ superlattice with compressive strain (c/a = 1.05) in the C-type magnetic configuration. The symbols $\uparrow$ and $\downarrow$ represent the spin majority and minority states with respect to a Mn atom. The spin-majority e$_g$ band splits into x$^2$-y$^2$ (unoccupied) and 3z$^2$-1 (partially occupied) bands as  seen more clearly from the band structure plot (Fig. \ref{band}).}
\end{figure}

 When the strain is compressive, the Mn-O bond length reduces in the MnO$_2$ plane while it increases between the planes. As a result, the 3z$^2$-1 orbital is lower in energy and is more occupied, while the x$^2$-y$^2$ orbital is higher in energy and is less occupied which is seen from the densities of states (Fig. \ref{doscaf}) as well as from the band structure (Fig. \ref{band}, right panel).
  
\begin{figure}
\includegraphics[width=6cm]{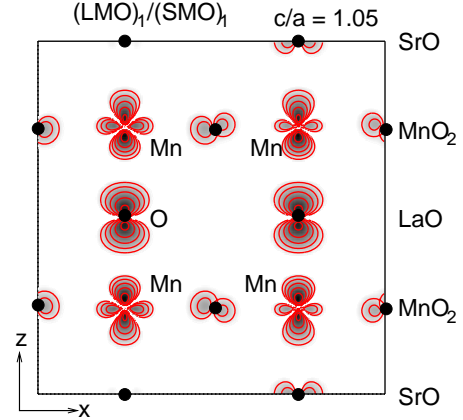}
\caption{\label{contourcaf} (color online). Valence electron charge-density contours plotted
on the xz-plane in the energy range E$_F$ - 0.15 eV to E$_F$ indicating the
orbital ordering for the C-type magnetic configuration. Contour values are: $\rho_n$ = $\rho_0 \times 10^{n\delta}$e/\AA$^3$, where $\rho_0$ = 3.7 $\times 10^{-3}$, $\delta$ = 0.4 and n labels the contours. The charge contours on the yz-plane (not shown) are identical to those plotted on the xz-plane. The orbital ordering is mainly 3z$^2$-1.}
\end{figure}

As in the case of A-type magnetic configuration in the tensile strain condition discussed earlier, here also 
we have two Mn$\uparrow$ and two Mn$\downarrow$ atoms.   So for the local spin majority channel, we have two x$^2$-y$^2$ orbitals and two 3z$^2$-1 orbitals. From Fig. \ref{band} we see that the x$^2$-y$^2$ orbitals lie in the conduction band and only one of the two 3z$^2$-1 orbitals lies in the valence band. This shows that the occupancy of the 3z$^2$-1 orbital per Mn atom is close to 0.5 and x$^2$-y$^2$ orbitals are basically unoccupied. This is seen from  the orbital occupancy (Fig. \ref{energyoccup}) as well as from the charge-density contour plot of Fig. \ref{contourcaf}, where we see that the orbital ordering is predominantly  Mn- 3z$^2$-1 and O-p$_z$. 

    The partially occupied 3z$^2$-1 orbital mediates  a strong double exchange mechanism to make the out-of-plane magnetic ordering ferromagnetic. The in-plane remains antiferomagnetic due to the superexchange between the localized t$_{2g}$ electrons.  Thus the net magnetic configuration for the compressive interface is C-type in agreement with our total energy calculations shown in Fig. \ref{energyoccup}.    
Experimental studies on (LMO)$_3$/(SMO)$_2$ interface, grown on
 LAO substrate (c/a = 1.05), do indeed show a C-type  antiferromagnetic configuration \cite{yamada} consistent with our theoretical results.
 
\begin{figure}
\includegraphics[width=5cm]{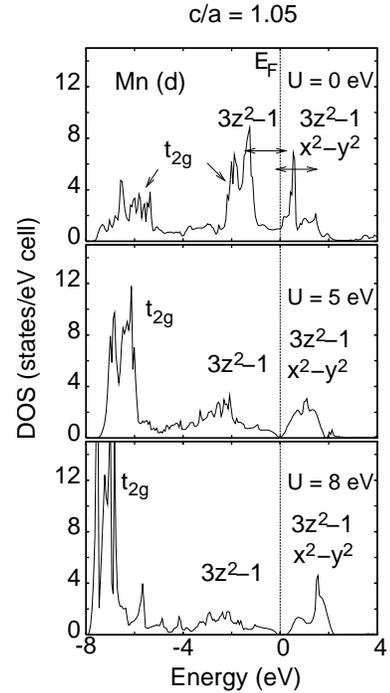}
\caption{\label{udoscaf} Atom projected Mn-d DOS for a single Mn atom in the spin-majority channel for the (LMO)$_1$/(SMO)$_1$ superlattice with the compressive strain condition (c/a = 1.05) and the C-type magnetic configuration. The results are obtained from GGA+U with different values of U. For U $\ge$ 5 eV the DOS has a gap at the Fermi level.
The $z^2-1$ state is split due to the interaction between the Mn atoms on adjacent planes, while the $x^2-y^2$ state is not, because of the AFM alignment of the neighboring Mn atoms in the plane.
}
\end{figure}

  Unlike the lattice matched interfaces which are metallic, the compressive interface is insulating. It is known that the strong correlation effect in manganites plays an important role to drive the insulating behavior. To elucidate this effect, in Fig. \ref{udoscaf} 
 we have plotted the Mn-d DOS for different values of U. For $U \ge 5$ eV, a gap opens at the Fermi level to make the interface insulating.

\section{Strain effect on magnetism: A three site model} 

So far, from the electronic structure calculations for  the (LMO)$_1$/(SMO)$_1$ superlattice under various strain conditions, we found that strain changes the relative occupancy of the two Mn-e$_g$ orbitals, which in turn affects the 
magnetic ordering in the structure. We found that the ordering is antiferromagnetic in the plane and ferromagnetic out of the plane, if the strain is compressive and opposite, if the strain is tensile, while  
 for the lattice-matched interface, the magnetic ordering is ferromagnetic in all directions. In this section we develop a simple three site model consisting of the Mn-O-Mn atoms to further understand the effect of the strain-controlled orbital occupancy on the magnetic interactions at the interface. 

   In bulk perovskite manganites, the t$_{2g}^3$ core spins interact via the  antiferromagnetic 
superexchange. In addition to this, the $e_g$ electrons mediate the ferromagnetic Anderson-Hasegawa double exchange between the core spins, which competes with the antiferromagnetic superexchange. The strength of the double exchange depends on which of the $e_g$ orbitals is occupied because of the anisotropic hopping and these are modeled in this Section.

   In the LMO/SMO superlattice, as seen from the density-functional results,
 the Mn atoms at the interface are left with one extra electron (0.5 electrons per Mn) which occupies the itinerant e$_g$ states. The epitaxial strain splits the degenerate Mn-e$_g$ states into x$^2$-y$^2$ and 3z$^2$-1 states. From the model below, we will see 
that depending on the strain condition the relative occupancies of the interface x$^2$-y$^2$ and 3z$^2$-1 orbitals could switch a ferromagnetic interaction into an antiferromagnetic one and vice versa.

\begin{figure}
\vspace{1cm}
\includegraphics[width=6cm]{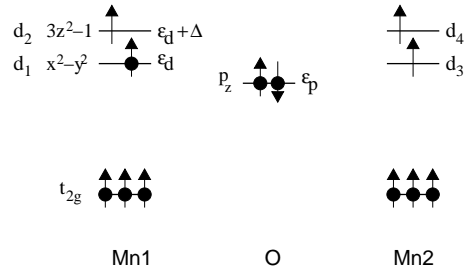}
\caption{\label{modelfm} Orbitals considered in forming the Hamiltonian
 H$_{\uparrow\uparrow}$ appropriate for the out of the plane ferromagnetic configuration at the LMO/SMO interface. The parameter $\Delta$ is the onsite energy difference
between the x$^2$-y$^2$ and 3z$^2$-1 orbitals. Depending on the strain condition $\Delta$ can be negative
(compressive strain) or positive (tensile strain).
The black dots indicate the occupied orbitals for one many-particle configuration, while the arrows indicate the spin states of the orbitals.}
\end{figure}

   In our model we consider the t$_{2g}$ electrons as classical core spins which are fixed at the Mn sites with negligible 
 intersite hopping as compared to the itinerant e$_g$ electrons, as has been used by many authors in the literature. Hence the Mn-O-Mn double exchange is due to the hopping between the itinerant x$^2$-y$^2$ and 3z$^2$-1 electrons and O-p electrons (Fig. \ref{modelfm}). The model Hamiltonian thus reads: 
\begin{eqnarray}
\nonumber H=\sum_{i\alpha\sigma}\epsilon_{i\alpha}n_{i\alpha\sigma}+\sum_{\langle ij \rangle \alpha \beta\sigma}t_{i\alpha j\beta} (c^{\dagger}_{i\alpha\sigma}c_{j\beta\sigma}+H.c.)\\
\nonumber +\frac{1}{2}\sum_{i} U_i n_i(n_i -1) - J_H\sum_{i\alpha}^{Mn1, Mn2}\overrightarrow{S}_{i}\cdot\overrightarrow{s}_{i\alpha}\\
 + \frac{1}{2}J_{SX}\overrightarrow{S}_{Mn1}\cdot\overrightarrow{S}_{Mn2}.
\label{Hamiltonian}
\end{eqnarray}
Here, i, $\alpha$ and $\sigma$ are, respectively, the site (Mn or O), orbital
(Mn- x$^2$-y$^2$, 3z$^2$-1, O-p$_x$, p$_y$, p$_z$) and spin indices. The parameter $\epsilon_{i\alpha}$ is the onsite energy of the orbital,
$\langle ij \rangle$ indicates nearest neighbors, c$^{\dagger}$s are the creation operators, and n$_i$ is the total
number of electrons at i-th site. 
The matrix elements  t$_{i\alpha j\beta}$ are the Slater-Koster tight binding hopping integrals between the Mn-e$_g$ and O-p
 orbitals.
We shall, for simplicity, take the Hund's coupling  J$_H$ as $\infty$ so that only the e$_g$ states parallel to the t$_{2g}$
spin at a Mn site can be occupied. The symbol $\vec{S}_i$ represents the t$_{2g}$ core spins and  $\vec{s}_{i\alpha}$ is the spin of the e$_g$ electron. The parameter J$_{SX}$ represents the superexchange between the t$_{2g}$ core spins. Throughout this paper, we have taken $U_p = 0$ and $U_d = 5$ eV unless otherwise stated and also the on-site energy of the oxygen orbitals are taken as zero: $\epsilon_p = 0$ for all spins and all three p orbitals.

\begin{table}
\begin{center}
\begin{tabular}{|c|c|c|c|c|}
\hline
Mn$\rightarrow$O&orbitals&$|p_x\rangle$&$|p_y\rangle$&$|p_z\rangle$\\
Direction&&&&\\
\hline
$\hat{x}$&$\langle x^2-y^2|$&-$\sqrt3$V$_{pd\sigma}/2$&0&0\\

&$\langle 3z^2-1|$& V$_{pd\sigma}/2$&0&0\\
\hline
$\hat{y}$&$\langle x^2-y^2|$&0&$\sqrt3$V$_{pd\sigma}/2$&0\\

&$\langle 3z^2-1|$&0&V$_{pd\sigma}/2$&0\\
\hline
$\hat{z}$&$\langle x^2-y^2|$&0&0&0\\

&$\langle 3z^2-1|$&0&0&-V$_{pd\sigma}$\\
\hline
\end{tabular}
\caption{Slater-Koster tight binding hopping integrals between the Mn-e$_g$ and O-p orbitals. In the three site model the value of V$_{pd\sigma}$ is taken as 0.9 eV.}
\end{center}
\end{table}

  The net exchange interaction between two Mn atoms is a sum of the antiferromagnetic superexchange J$_{SX}$, modeled by the last term in the model Hamiltonian Eq. (\ref{Hamiltonian}), and the double exchange J$_{DX}$  mediated by the itinerant e$_g$ electrons, modeled by the remaining terms
  in the same equation, so that we have
  \begin{equation}
  J=J_{SX}+J_{DX}.
  \end{equation}
The exchange interaction J is obtained by
 calculating the difference between the  ground state energies corresponding to the ferromagnetic (FM) and the antiferromagnetic (AFM) alignment of the two t$_{2g}$ core spins:
\begin{equation}
J =  E_{\uparrow\uparrow}-E_{\uparrow\downarrow}.
\end{equation}
Note that a positive (negative) value of J indicates an AFM (FM) interaction.
J$_{SX}$ is a simple additive term and it is, for the manganites, of the order of 26 meV \cite{hakim}. From the present model
we will calculate the J$_{DX}$ which depends on the occupancy of the e$_g$ states. 

   Before we move on to the solution of the model, there is another point that needs to be made. The magnetic interaction between the planes, which is indicated by J in Fig. \ref{magfig} may differ when we consider the exchange interaction between the Mn spins via the O-p orbitals across the LaO plane or the SrO plane. However, from the charge contours ({\it e.g.} Fig. \ref{contouraaf}), we see that there is very little difference between the two oxygen atoms located on these planes, so that the  Mn-O-Mn coupling may be expected to be nearly the same. So, in our model, we do not differentiate between these two interactions, so that J is the same across the LaO plane or the SrO plane.
To distinguish the exchange interaction out of the plane and in the plane, we have used the notation J$_{DX}$ for the former and J$_{DX}^{\prime}$ for the latter in the remaining part of this section.

\subsection{Out-of-plane exchange J}

 First consider the out-of-plane exchange by evaluating the ground state energies for the ferromagnetic and the
antiferromagnetic configurations of the Mn-t$_{2g}$ spins of two Mn atoms along the z-axis.  The model for the out-of-plane FM configuration is schematically shown in Fig. \ref{modelfm}. Listed in Table I are the Slater-Koster tight binding hopping integrals between Mn-e$_g$ and O-p orbitals, which shows that out of the three O-p orbitals, only p$_z$ takes part in the hopping process along the z-axis. Therefore, in this model we have five spin-up
orbitals (O-p$_z$, two Mn-z$^2$-1 and two Mn- x$^2$-y$^2$) available for two spin-up electrons
and one spin-down orbital (O-p$_z$) available for the lone spin-down electron.
The spin-down electron can only be on the O-p$_z$ orbital in the ferromagnetic case due to the infinite J$_H$ and does not
take part in the hopping process. Hence, we have a ten dimensional
two particle configuration space ($^5C_2 \times ^1C_1$). 

   We choose the two-particle (both particles with spins up) basis set in the order:
 $\vert pd_1\rangle$, $\vert d_1d_2\rangle$, $\vert d_1d_4\rangle$,
 $\vert pd_2\rangle$, $\vert pd_4\rangle$, $\vert d_2d_4\rangle$,
 $\vert pd_3\rangle$, $\vert d_2d_3\rangle$, $\vert d_3d_4\rangle$, and
$\vert d_1d_3\rangle$,  where p, d$_1$, d$_2$, d$_3$ and d$_4$ respectively denote the O-p$_z$, Mn1-x$^2$-y$^2$, Mn1-3z$^2$-1, Mn2-x$^2$-y$^2$, and Mn2-3z$^2$-1 orbitals. With this basis set, for the case of the two Mn-t$_{2g}$ core spins ferromagnetically aligned, the Hamiltonian for the itinerant electrons H$_{\uparrow\uparrow}$ becomes: 
\begin{widetext}
\begin{eqnarray}
H_{\uparrow\uparrow}
=\left( \begin{array}{cccc}
\left[\begin{array}{ccc}
\epsilon_d&t&-t\\
t& 2\epsilon_d+\Delta+U_d&0\\
-t&0&2\epsilon_d+\Delta\\
\end{array}\right] &&0&\\
&\left[\begin{array}{ccc}
\epsilon_d+\Delta&0&-t\\
0&\epsilon_d+\Delta&-t\\
-t&t&2\epsilon_d+2\Delta\\
\end{array}\right] &&\\
&&\left[\begin{array}{ccc}
\epsilon_d&t&-t\\
t&2\epsilon_d+\Delta&0\\
-t&0&2\epsilon_d+\Delta+U_d\\
\end{array}\right]&\\
&0&&\left[\begin{array}{c}
2\epsilon_d\\
\end{array}\right]\\
\end{array}
\right)
.
\label{hofm}
\end{eqnarray}
%\end{@twocolumnfalse}
%]
%\makeatother
\end{widetext}

  Again, here $\epsilon_d$ represents the 
onsite energy for the Mn-x$^2$-y$^2$ orbitals and $\Delta$ is the energy shift of the Mn-z$^2$-1 orbital from the x$^2$-y$^2$ orbital due to strain. The onsite energy of the O-p orbitals is taken as zero. Parameter t is the hopping matrix element (V$_{pd\sigma}$). From DFT calculations, we found that for compressive strain condition $\Delta$ is +ve and for
tensile strain condition $\Delta$ is -ve. For the lattice matched interface the onsite energy of the x$^2$-y$^2$ and the 3z$^2$-1 orbitals are about the same.

\begin{figure}
\includegraphics[width=6cm]{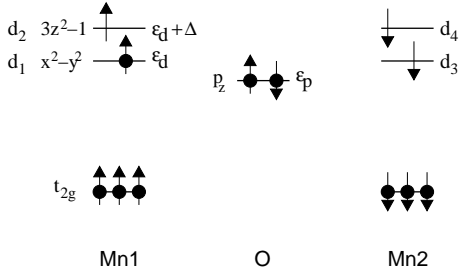}
\caption{\label{modelafm} Orbitals considered in forming the Hamiltonian
 H$_{\uparrow\downarrow}$ appropriate for the out-of-plane exchange.
 Of the three oxygen p orbitals, only p$_z$ has a non-zero hopping to the 
 Mn e$_g$ orbitals.}
\end{figure}

  We now consider the antiferromagnetic case, where the two Mn-t$_{2g}$ spins are aligned antiferromagnetically. 
In this case, as seen from Fig. \ref{modelafm}, we see that there are six active
orbitals, three spin-up and three spin-down. One can populate these orbitals with two spin-up electrons and one spin-down
electrons or vice versa. 
(We do not consider the configurations where all three electrons have the same spins as this would correspond to an oxygen-to-manganese charge transfer state, which has a much higher energy.)
In either of these cases we have a nine dimensional three particle configuration space ($^3C_2\times^3C_1$).  These two sets of configurations do not interact with each 
other,  as the model Hamiltonian Eq. (\ref{Hamiltonian}) does not allow
hopping between two opposite spins, and have the same ground-state energy.

  By considering the configurations with two spin-up electrons and one spin-down electron and arranging the basis set in the order
$\vert p\bar{p}d_1\rangle$, $\vert \bar{p}d_1d_2\rangle$, $\vert pd_1\bar{d_4}\rangle$, $\vert d_1d_2\bar{d_4}\rangle$, $\vert pd_1\bar{d_3}\rangle$, $\vert d_1d_2\bar{d_3}\rangle$, $\vert p\bar{p}d_2\rangle$, $\vert pd_2\bar{d_4}\rangle$,
 and $\vert pd_2\bar{d_3}\rangle$, where the bar stands for the spin-down orbitals and unbar stands for the spin-up orbitals, the antiferromagnetic Hamiltonian, H$_{\uparrow\downarrow}$, becomes:
\begin{widetext}
\begin{eqnarray}
H_{\uparrow\downarrow}
=\left( \begin{array}{cccc}
\left[\begin{array}{cccc}
\epsilon_d&t&t&0\\
t& 2\epsilon_d+\Delta+U_d&0&t\\
t&0&2\epsilon_d+\Delta&t\\
0&t&t&3\epsilon_d+2\Delta+U_d\\
\end{array}\right]&&0&\\
&\left[\begin{array}{cc}
2\epsilon_d&t\\
t&3\epsilon_d+\Delta+U_d\\
\end{array}\right] &&\\
&&\left[\begin{array}{cc}
\epsilon_d+\Delta&t\\
t&2\epsilon_d+2\Delta\\
\end{array}\right]&\\
&0&&\left[\begin{array}{c}
2\epsilon_d+\Delta\\
\end{array}\right]\\
\end{array}
\right)
.
\label{hoafm}
\end{eqnarray}
%\end{@twocolumnfalse}
%]
%\makeatother
\end{widetext}

The above matrices can be diagonalized easily for all parameter values; however, for some limiting cases, one can solve these either analytically or using the perturbation theory, which then gives us considerable insights into the resulting exchange interactions.

To this end, we first consider the limit when 
 $\Delta$ is sufficiently large and negative. One immediately sees by inspecting the matrix that the ground state of $H_{\uparrow\uparrow}$ comes from the second block diagonal
of Eq. (\ref{hofm}). Diagonalization then yields the ground state energy to be
\begin{equation} 
E_{\uparrow\uparrow} = 3(\epsilon_d+\Delta)/2- \sqrt{(\epsilon_d+\Delta)^2+8t^2}/2.
\label{eofmcompressive}
\end{equation} 
For the AFM Hamiltonian $H_{\uparrow\downarrow}$ in the same limit for $\Delta$, the ground state comes from the third block diagonal of Eq. (\ref{hoafm}), yielding immediately the ground-state energy 
\begin{equation}
E_{\uparrow\downarrow} = 3(\epsilon_d+\Delta)/2 - \sqrt{(\epsilon_d+\Delta)^2+4t^2} /2.
\label{eoafmcompressive}
\end{equation}

\begin{figure}
\includegraphics[width=8cm]{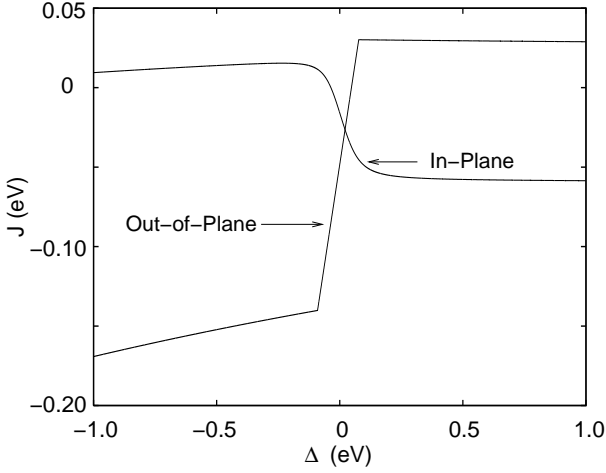}
\caption{\label{jexch} Exchange parameter J (= J$_{DX}$ + J$_{SX}$), with J$_{DX}$ obtained from the numerical diagonalization of the Hamiltonians (Eq. (\ref{hofm}), (\ref{hoafm}), (\ref{hifm}), and (\ref{hiafm})), as a function of
 energy difference $\Delta$ between the x$^2$-y$^2$ and the 3z$^2$-1 orbitals. For the calculation of J, the parameters $\epsilon_d$, U$_d$, V$_{pd\sigma}$ and J$_{SX}$ are taken as 5 eV, 5 eV, 0.9 eV, and 26 meV respectively. When $\Delta$ is sufficiently positive
(tensile strain) the magnetic interaction is FM in the plane ($J^\prime$) and AFM between the planes (J).
When $\Delta$ is sufficiently negative (compressive strain) the magnetic interaction reverses.}
\end{figure}

  The out-of-plane exchange energy J$_{DX}$ is obtained by taking the difference between the FM and AFM ground state energies, so that for the case of the large and negative values of $\Delta$ that we are considering,  we get
  \begin{eqnarray}
\nonumber J_{DX} &=& E_{\uparrow\uparrow}-E_{\uparrow\downarrow} \\
\nonumber &=& 1/2 \times (\sqrt{(\epsilon_d+\Delta)^2+4t^2}-\sqrt{(\epsilon_d+\Delta)^2+8t^2}).\\
\label{jocompressive}
  \end{eqnarray}
 As seen from Eq. (\ref{jocompressive}), the exchange energy is negative  irrespective of the Hamiltonian parameters, indicating that the out-of-plane double exchange is ferromagnetic and robust. We have to add the superexchange term $J_{SX}$ to this to get the net magnetic interaction. The results are consistent with the numerically computed value of J presented in Fig. \ref{jexch}, which is discussed later.

Similarly, we can also obtain an expression for a large and positive $\Delta$. Again, it is immediately clear from the inspection of the Hamiltonians Eqs. (\ref{hofm}) and (\ref{hoafm}) that the FM and AFM ground states come from the first sub-block of H$_{\uparrow\uparrow}$ and H$_{\uparrow\downarrow}$. An analytical diagonalization is not possible in this case, however one can
 apply the fourth order non-degenerate perturbation theory to
 compute  $E_{\uparrow\uparrow}$ and $E_{\uparrow\downarrow}$, the difference of which yields the result
\begin{equation}
J_{DX} = \frac{t^4}{2\epsilon_d+2\Delta+U_d}
( \frac{1}{\epsilon_d+\Delta+U_d}+\frac{1}{\epsilon_d+\Delta})^2.
\label{jotensile}
\end{equation}
We see that if the x$^2$-y$^2$ orbital is more occupied (i. e. $\Delta$ $>$ 0), the out-of-plane  magnetic ordering is AFM for any values of $\epsilon_d$ and $\Delta$.

In Fig. \ref{jexch}, we have calculated the exchange interaction by simply a numerical diagonalization of the two Hamiltonian matrices for a general value of $\Delta$, viz., -1.0 eV $\le$ $\Delta$ $\le$ 1.0 eV. The numerical results for J are consistent with with our analytical and perturbation results for the large values of $\Delta$. When $\Delta$ is less than -0.15 eV we have a FM interaction and if it is greater than 0.15 eV we have an AFM interaction. For the intermediate region of $\Delta$ (-0.15 eV $\le$ $\Delta$ $\le$ 0.15 eV), the ground state swiftly changes between FM and AFM. 

    Our model is consistent with the DFT calculations discussed in the previous section (Fig. \ref{energyoccup}), where we see that for compressive strain the 
 3z$^2$-1 orbitals are more occupied ($\Delta<$ 0) and the total energy calculation yields a out-of-plane FM configuration. If the strain
is tensile, the x$^2$-y$^2$ orbital is more occupied ($\Delta>$ 0)and the out-of-plane magnetic ordering is AFM.

\subsection{In-plane exchange J$^\prime$}

Now we consider the Mn-O-Mn exchange in the plane (xy-plane), i.e., between two Mn atoms on a MnO$_2$ plane adjacent to the interface. In this 
case p$_x$, p$_y$ and e$_g$ are the only active orbitals since 
the hopping between p$_z$ and e$_g$ orbitals is not allowed (Table I). For concreteness we have taken the Mn-O-Mn to be along the x-axis and results are identical if it is along the y-axis. 

First consider the FM configuration of the two Mn-t$_{2g}$ spins. Here the itinerant electrons have a 10 dimensional configuration space as in the case of out-of-plane, but with p$_x$ as the active orbital instead of p$_z$.  Taking the two-particle basis set in the 
order 
$\vert pd_1\rangle$, $\vert pd_2\rangle$, $\vert pd_3\rangle$, $\vert pd_4\rangle$, $\vert d_1d_2\rangle$, 
$\vert d_1d_3\rangle$, $\vert d_1d_4\rangle$, $\vert d_2d_3\rangle$, $\vert d_2d_4\rangle$, and $\vert d_3d_4\rangle$ and denoting by t$^{\prime}$ and t$^{\prime\prime}$ the hopping matrix elements $\frac{V_{pd\sigma}}{2}$ and
$\frac{\sqrt3V_{pd\sigma}}{2}$, respectively,
 the Hamiltonian H$_{\uparrow\uparrow}$ becomes
\begin{widetext}
\begin{eqnarray}
H_{\uparrow\uparrow}=\left(
\begin{array}{cccccccccc}
\epsilon_d&0&0&0&-t^{\prime}&-t^{\prime\prime}&t^{\prime}&0&0&0\\
0& \epsilon_d+\Delta&0&0&-t^{\prime\prime}&0&0&-t^{\prime\prime}&t^{\prime}&0\\
0&0&\epsilon_d&0&0&-t^{\prime\prime}&0&t^{\prime}&0&t^{\prime}\\
0&0&0&\epsilon_d+\Delta&0&0&-t^{\prime\prime}&0&t^{\prime}&t^{\prime\prime}\\
-t^{\prime}&-t^{\prime\prime}&0&0&2\epsilon_d+\Delta+U_d&0&0&0&0&0\\
-t^{\prime\prime}&0&-t^{\prime\prime}&0&0&2\epsilon_d&0&0&0&0\\
t^{\prime}&0&0&-t^{\prime\prime}&0&0&2\epsilon_d+\Delta&0&0&0\\
0&-t^{\prime\prime}&t^{\prime}&0&0&0&0&2\epsilon_d+\Delta&0&0\\
0&t^{\prime}&0&t^{\prime}&0&0&0&0&2\epsilon_d+2\Delta&0\\
0&0&t^{\prime}&t^{\prime\prime}&0&0&0&0&0&2\epsilon_d+\Delta+U_d\\
\end{array}
\right)
.
\label{hifm}
\end{eqnarray}
\end{widetext}

  Similar to out-of-plane AFM Hamiltonian, the in-plane AFM Hamiltonian has a nine dimensional configuration space. If we choose the basis set in the order
$\vert p\bar{p}d_1\rangle$,
 $\vert p\bar{p}d_2\rangle$,
 $\vert \bar{p}d_1d_2\rangle$,
 $\vert pd_1\bar{d_3}\rangle$,
 $\vert pd_2\bar{d_3}\rangle$,
 $\vert d_1d_2\bar{d_3}\rangle$,
 $\vert pd_1\bar{d_4}\rangle$,
 $\vert pd_2\bar{d_4}\rangle$, and $\vert d_1d_2\bar{d_4}\rangle$, the antiferromagnetic Hamiltonian  $H_{\uparrow\downarrow}$ for the present case reads
 %
%\end{@twocolumnfalse}
%]
%\makeatother
\begin{widetext}
\begin{equation}
H_{\uparrow\downarrow}=\left(
\begin{array}{ccccccccc}
\epsilon_d&0&-t^{\prime}&t^{\prime\prime}&0&0&-t^{\prime}&0&0\\
0& \epsilon_d+\Delta&-t^{\prime\prime}&0&t^{\prime\prime}&0&0&-t^{\prime}&0\\
-t^{\prime}&-t^{\prime\prime}&2\epsilon_d+\Delta+U_d&0&0&t^{\prime\prime}&0&0&-t^{\prime}\\
t^{\prime\prime}&0&0&2\epsilon_d&0&-t^{\prime}&0&0&0\\
0&t^{\prime\prime}&0&0&2\epsilon_d+\Delta&-t^{\prime\prime}&0&0&0\\
0&0&t^{\prime\prime}&-t^{\prime}&-t^{\prime\prime}&3\epsilon_d+\Delta+U_d&0&0&0\\
-t^{\prime}&0&0&0&0&0&2\epsilon_d+\Delta&0&-t^{\prime}\\
0&-t^{\prime}&0&0&0&0&0&2\epsilon_d+2\Delta&-t^{\prime\prime}\\
0&0&-t^{\prime}&0&0&0&-t^{\prime}&-t^{\prime\prime}&3\epsilon_d+2\Delta+U_d\\
\end{array}
\right)
.
\label{hiafm}
\end{equation}
%\end{@twocolumnfalse}
%]
%\makeatother
\end{widetext}

  An analytical diagonalization for both the ferromagnetic and antiferromagnetic Hamiltonian  is not possible due to their
non block diagonal nature and large dimension. Non-degenerate perturbation theory can not be applied as H$_{\uparrow\uparrow}$ contains degenerate states. 
However, in the limit $\Delta \rightarrow \infty$ the ground state energies for the FM and AFM configuration is obtained from a
3$\times$3 sub-matrix of H$_{\uparrow\uparrow}$ and a 2$\times$2 sub-matrix of H$_{\uparrow\downarrow}$ respectively. The in-plane exchange energy J$_{DX}^{\prime}$ in the limit $\Delta \rightarrow \infty$ is then
immediately obtained as
\begin{equation}
J_{DX}^{\prime} = (\sqrt{\epsilon_d^2+4t^{\prime\prime 2}} - \sqrt{\epsilon_d^2+8t^{\prime\prime 2}}       )/2.
\label{jitensile}
\end{equation}
 Here, we see that for large and positive value of $\Delta$, J$_{DX}$ is  negative quantity which implies a FM interaction in the plane.

  Similarly, in the limit $\Delta \rightarrow -\infty$, we find that
\begin{equation}
J_{DX}^{\prime} = [\epsilon_d+U_d+\sqrt{(\epsilon_d+U_d)^2+4t^{\prime\prime 2}}]/2.
\label{jicompressive}
\end{equation}
 Eq. (\ref{jicompressive}) shows that for large and negative value of $\Delta$, we have an AFM configuration in the plane. 
 
 In Fig. \ref{jexch} we have plotted
 the in-plane exchange J$^{\prime}$ as a function of $\Delta$ obtained from direct numerical diagonalization of the full Hamiltonians H$_{\uparrow\uparrow}$ and H$_{\uparrow\downarrow}$, which indeed shows that for the in-plane exchange J$^\prime$, the magnetic interaction in the plane switches from ferromagnetic to antiferromagnetic, as $\Delta$ is changed from   positive to
negative values.  
\begin{table}
\begin{center}
\begin{tabular}{c|cc|ccc}
\hline
& DFT &&& Model  & \\
c/a&J$^{\prime}$ &J &$\Delta$ (eV)&J$^{\prime}$  &J \\
\hline
0.95&-85&34&1.0&-58&29\\
1.00&-27&-30&0.0&-18&-44\\
1.05&14&-100&-1.0&11&-164\\
\hline
\end{tabular}
\caption{Mn-O-Mn exchange energy (in meV) in the plane (J$^{\prime}$) and out of the plane (J) obtained from the 
DFT calculations and the three-site model for different values of the c/a ratio.
 The model uses the $\Delta$ values corresponding roughly to the DFT results 
 for the three different strain conditions and the other parameters are: $\epsilon_d = U_d = 5$ eV, V$_{pd\sigma}=0.9$ eV, and 
 J$_{SX}$ = 26 meV.\cite{hakim}  
 }
\end{center}
\end{table}

To summarize our analysis in the Section, we find that the strain-induced splitting in the e$_g$ states is instrumental in determining the magnetic properties at the interface. If x$^2$-y$^2$ orbital is relatively more occupied than the 3z$^2$-1 orbital, the magnetic ordering  at the LMO/SMO interface is more likely to be A-type with the FM and AFM configurations stabilized in the pane and out of the plane respectively. If the 3z$^2$-1 orbital is more
occupied the magnetic ordering is more likely to be C-type with in-plane AFM ordering and out-of-plane FM ordering. When both the e$_g$ orbitals are more or less equally occupied, the double exchange wins over the superexchange to stabilize the interface
in a three dimensional FM configuration.

 The exchange interactions calculated from the three site model are in good agreement with the results obtained from DFT calculations as shown inTable II.
 
\section{Effect of strain on a thicker superlattice: ($LaMnO_3$)$_1$/($SrMnO_3$)$_3$}

  In the preceding sections we have studied the effects of strain on the magnetic interactions in the  
(LMO)$_1$/(SMO)$_1$ superlattice.  To generalize the strain effects on magnetism at the interface,  in this section we have analyzed the magnetic properties of a thicker superlattice, viz., (LMO)$_1$/(SMO)$_3$, which has both inner and interfacial MnO$_2$ planes.

\begin{figure}
\includegraphics[width=5.5cm]{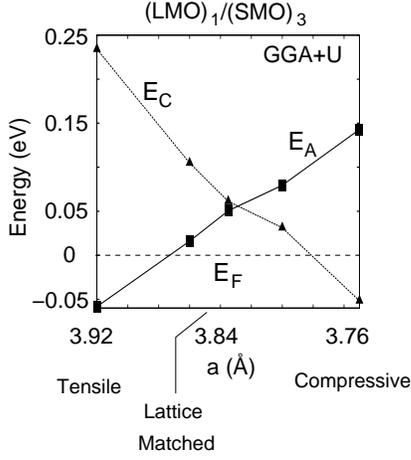}
\caption{\label{energy13}Total energies for magnetic configurations A and C  relative to the energy for the magnetic configuration F, as a function of the in-plane lattice parameter `a' for the (LMO)$_1$/(SMO)$_3$ superlattice. The magnetic configurations A, F, and C are shown in Fig. \ref{magfig}. For strong tensile strain condition, we see a stable A-type magnetic ordering for the interfacial MnO$_2$ layers, while for strong compressive strain condition the magnetic ordering is C-type. For the lattice matched case, the ferromagnetic ordering is stabilized in all directions.}
\end{figure}

   In the (LMO)$_1$/(SMO)$_3$ superlattice, we have taken the G-type AFM
configuration for the inner MnO$_2$ planes as they belong to the SMO constituent of the superlattice. For the interfacial MnO$_2$ planes, we have considered again the F-type, A-type and C-type magnetic configurations (Fig. \ref{magfig}). 
 In Fig. \ref{energy13}, we have shown the energetics for these three magnetic configurations as a function of the in-plane lattice parameter `a'. From the figure we see that for lower values of the lattice constant `a', the interface shows a stable
C-type magnetic ordering and as we increase the value of `a', the interface gradually moves towards an F-type magnetic configuration. For very high values of `a', the interface stabilizes with the A-type magnetic configuration.

  For a strain free LMO/SMO superlattice, the in-plane lattice parameter coincides
with the average lattice parameter a$_0$ \cite{yamada} (see section-II), which is 3.835 {\AA} for
the (LMO)$_1$/(SMO)$_3$ superlattice. 
As in the case of (LMO)$_1$/(SMO)$_1$ 
superlattice (Fig. \ref{energyoccup}), here also we see that for strong compressive strain condition, the stable magnetic ordering at the interface is C-type, while for strong tensile strain condition the magnetic ordering is F-type. For the lattice 
matched interface we see a stable FM ordering (F-type) in all directions. Our band structure calculation shows insulating behavior for strong compressive strain condition, while for lattice matched and tensile strain conditions it is metallic. 

\begin{figure}
\includegraphics[width=6cm]{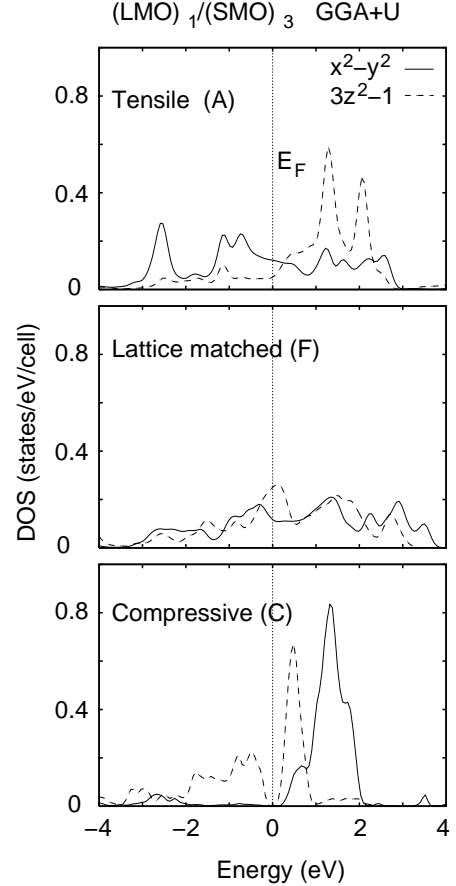}
\caption{\label{dos13}Spin majority Mn-e$_g$ DOS corresponding to the interface Mn atoms for the (LMO)$_1$/(SMO)$_3$ superlattice. The results are shown for the in-plane lattice parameter `a' = 3.92 {\AA} (tensile), 3.835 {\AA} (lattice matched) and 3.765 {\AA} (compressive). If the strain condition is tensile, the x$^2$-y$^2$ orbital is more occupied and the 3z$^2$-1 orbital is less occupied, and opposite if the strain condition is compressive. For the lattice matched condition, both  x$^2$-y$^2$ and 3z$^2$-1 orbitals are more or less equally occupied. 
%The lattice matched interface shows half-metallic behavior.
}
\end{figure}

  To study the role of Mn-e$_g$ orbitals on the interfacial magnetism in  (LMO)$_1$/(SMO)$_3$ superlattice, in Fig. \ref{dos13} we have shown the orbital projected Mn-e$_g$ DOS for the interface Mn atoms for a = 3.92 {\AA} (tensile strain), 3.835 {\AA} (lattice matched) and 3.75 {\AA} (compressive strain). From the figure we see that, analogous to the case of (LMO)$_1$/(SMO)$_1$ superlattice, for the tensile strain condition the x$^2$-y$^2$ is relatively more occupied and 3z$^2$-1 orbital is less occupied, and opposite for the compressive strain condition. For the lattice matched interface, both the x$^2$-y$^2$ and 3z$^2$-1 orbitals are more or less
equally occupied as before. 

   Hence, as discussed earlier, for the tensile strain condition, we have a strong ferromagnetic double exchange coupling in the MnO$_2$ plane through x$^2$-y$^2$ orbitals while between the planes we have a antiferromagnetic coupling due to superexchange between the t$_{2g}$ core spins. For the compressive strain condition the higher occupancy of Mn-3z$^2$-1 orbitals leads to a ferromagnetic coupling between the planes and due to negligible occupancy of the x$^2$-y$^2$ orbital, the antiferromagnetic ordering is sustained in the MnO$_2$ plane. For the lattice matched structure, the orbital ordering is a combination of both x$^2$-y$^2$ and 3z$^2$-1 orbitals and hence strong double exchange coupling both in the plane and out of the plane stabilizes the ferromagnetic interaction in all directions. The similarity in the interfacial magnetic properties for both the superlattices, viz., (LMO)$_1$/(SMO)$_1$ and (LMO)$_1$/(SMO)$_3$ for different strain conditions  suggests that the strain effect on  magnetism at the LMO/SMO interface may be true
   for the general (LMO)$_n$/(SMO)$_m$ superlattice as well. 

\section{Summary}

In summary, we have studied the effect of the epitaxial strain on the magnetic ordering at the interface of LMO/SMO superlattices by a detailed analysis on the (LMO)$_1$/(SMO)$_1$ superlattice. We found that the epitaxial strain induces different orbital ordering which in turn changes the magnetic ordering at the interface. The magnetic ordering at the interface is determined by the competition between the antiferromagnetic
 superexchange between the core t$_{2g}$ electrons and ferromagnetic double exchange between the itinerant e$_g$ electrons. The strength of the latter in the MnO$_2$ plane or between the planes strongly depends on the occupancy of the non-degenerate e$_g$ orbitals.

 In case of a strong tensile strain condition, the higher occupancy of the x$^2$-y$^2$ orbital strengthens the 
double exchange coupling to stabilize the ferromagnetic ordering in the MnO$_2$ plane, while between the 
MnO$_2$ planes the reduced double exchange coupling, due to the negligibly occupied 3z$^2$-1 orbital, fails to 
overcome the antiferromagnetic t$_{2g}$-t$_{2g}$ superexchange and stabilizes the A-type magnetic ordering at the 
interface.  For strong compressive strain condition, the magnetic ordering reverses, 
viz., higher occupancy of the 3z$^2$-1 orbital lead to a ferromagnetic coupling between the MnO$_2$ planes, while
the depleted x$^2$-y$^2$ occupancy allows the antiferromagnetic ordering in the plane to make the C-type magnetic 
configuration as the most stable one. For a lattice matched structure, double exchange is strong enough both in the 
MnO$_2$ plane and between the planes, due to more or less equally occupied x$^2$-y$^2$ and 3z$^2$-1 orbitals, to allow 
ferromagnetic ordering in all directions. 

  The electronic structure calculations for the (LMO)$_1$/(SMO)$_3$ superlattice showed that the epitaxial strain affects the magnetism at the interface in a similar way as  the (LMO)$_1$/(SMO)$_1$ superlattice.  This suggests that the strain effect on magnetism may be similar for the general (LMO)$_n$/(SMO)$_m$ superlattice. In addition, similar considerations regarding the effects of strain on orbital ordering and magnetism should be valid 
for interfaces between other perovskite oxides as well.

  This work was supported by the U.S. Department of Energy under Grant No. 
DE-FG02-00ER45818.

\end{document}